\documentclass[12pt]{article}

\begin{document}

\begin{titlepage}

\begin{flushright}
IUHET-498\\
\end{flushright}
\vskip 2.5cm

\begin{center}
{\Large \bf Synchrotron and Inverse Compton Constraints on Lorentz Violations for
Electrons}
\end{center}

\vspace{1ex}

\begin{center}
{\large Brett Altschul\footnote{{\tt baltschu@indiana.edu}}}

\vspace{5mm}
{\sl Department of Physics} \\
{\sl Indiana University} \\
{\sl Bloomington, IN 47405 USA} \\

\end{center}

\vspace{2.5ex}

\medskip

\centerline {\bf Abstract}

\bigskip

We present a method for constraining Lorentz violation in the electron sector,
based on observations of the photons emitted by high-energy astrophysical sources.
The most important Lorentz-violating operators at the relevant energies are
parameterized by a tensor $c^{\nu\mu}$ with nine independent components. If $c$ is
nonvanishing, then there may be either a maximum electron velocity less than the
speed of light or a maximum energy for subluminal electrons; both
these quantities will
generally depend on the direction of an electron's motion. From synchrotron
radiation, we may infer a lower bound on the maximum velocity, and from inverse
Compton emission, a lower bound on the maximum subluminal energy. With observational
data for both these types of emission from multiple celestial sources, we may then
place bounds on all nine of the coefficients that make up $c$.
The most stringent bound, on a certain combination of the coefficients, is at the
$6\times 10^{-20}$ level, and bounds on the coefficients individually range from
the $7\times 10^{-15}$ level to the $2\times 10^{-17}$ level. For most of the
coefficients, these are the most precise bounds available, and with newly available
data, we can already improve over previous bounds obtained by the same
methods.

\bigskip

\end{titlepage}

\newpage

\section{Introduction}

Currently, there is a great deal of interest in the possibility that Lorentz and
CPT symmetries might not be exact in nature. If violations of these very basic
symmetries are discovered, they would be of tremendous importance. They would be
a very important clues about the nature of physics at the Planck scale. Many
candidate theories of quantum gravity suggest the possibility of Lorentz symmetry
breaking in certain regimes. For example, Lorentz violation could arise
spontaneously in
string theory~\cite{ref-kost18,ref-kost19} or elsewhere~\cite{ref-altschul7}.
There could also be Lorentz-violating 
physics in loop quantum gravity~\cite{ref-gambini,ref-alfaro} and
non-commutative
geometry~\cite{ref-mocioiu,ref-carroll3} theories, or Lorentz violation through
spacetime-varying couplings~\cite{ref-kost20}, or
anomalous breaking of Lorentz and CPT symmetries~\cite{ref-klinkhamer}
in certain spacetimes.

Ultimately, the correctness of Lorentz symmetry must be verified experimentally.
To date, there have been many high-precision experimental tests of Lorentz
invariance. These have included studies of matter-antimatter asymmetries for
trapped charged particles~\cite{ref-bluhm1,ref-bluhm2,ref-gabirelse,
ref-dehmelt1} and bound state systems~\cite{ref-bluhm3,ref-phillips},
determinations of muon properties~\cite{ref-kost8,ref-hughes}, analyses of
the behavior of spin-polarized matter~\cite{ref-kost9,ref-heckel2},
frequency standard comparisons~\cite{ref-berglund,ref-kost6,ref-bear,ref-wolf},
Michelson-Morley experiments with cryogenic resonators~\cite{ref-antonini,
ref-stanwix,ref-herrmann}, Doppler effect measurements~\cite{ref-saathoff,ref-lane1},
measurements of neutral meson
oscillations~\cite{ref-kost10,ref-kost7,ref-hsiung,ref-abe,ref-link,ref-aubert},
polarization measurements on the light from distant galaxies~\cite{ref-carroll1,
ref-carroll2,ref-kost11,ref-kost21}, and others.

There are many systems and reaction processes that could potentially be used to
set further bounds on Lorentz violation.
This work focuses on experimental limits based on observations of synchrotron
and inverse Compton (IC) radiation from ultrarelativistic electrons in astrophysical
sources. Some of these limits have already
appeared~\cite{ref-altschul5,ref-altschul6}, but with new experimental
data available, even better bounds are now possible.

In order to evaluate the results of sensitive Lorentz tests,
it has been useful to develop a local effective field theory that
parameterizes all possible Lorentz violations.
The most general such theory is the standard model extension
(SME)~\cite{ref-kost1,ref-kost2}. The SME includes, in addition to all
Lorentz-violating operators that can be written down using standard model fields,
all possible Lorentz-violating terms in the gravitational sector as
well~\cite{ref-kost12}.
Although many theories describing new physics suggest the possibility of Lorentz
violation, none of them are understood well enough to make firm predictions. The
utility of the SME is its generality. Working within the SME, we can place bounds on
the coefficients that parameterize any Lorentz violation, and these bounds do not
depend on the ultimate mechanism underlying the Lorentz violation.

Of course, the number of parameters in the most general version of the SME is
infinite. Practically, it is usually more useful to restrict attention to a finite
subset of these coefficients. The most commonly considered subset is the minimal
SME. This includes operators which are superficially renormalizable (that is, of
dimension three or four) and invariant under the standard model's $SU(3)_{c}\times
SU(2)_{L}\times U(1)_{Y}$ gauge group. The one-loop renormalizability of a
further subset of the minimal SME---the minimal QED extension---has been verified
explicitly~\cite{ref-kost4}. Moreover, radiative corrections to this
particular theory can
be extremely interesting. Such interesting effects as photon
splitting~\cite{ref-kost5}, ambiguous radiative
corrections~\cite{ref-jackiw1,ref-victoria1,ref-altschul1,ref-altschul2}, and
radiatively induced photon masses~\cite{ref-altschul8} are possible. The stability
of the minimal SME operators, as well as their causality properties, have also
been looked at in detail~\cite{ref-kost3}.

Many of the coefficients in the minimal
SME have been tightly constrained, but many others have not. For example, the
neutrino sector is barely constrained at all. This work presents the best bounds
on several coefficients in an important sector of the theory.

One of the radiation processes considered here---synchrotron radiation---has
already been the subject of a number of analyses that included Lorentz violations.
Several of these have focused on Lorentz violation through
changes to particle dispersion relations. This follows the popular approach of
Myers and Pospelov~\cite{ref-myers}. Taking a preferred direction
$v^{\mu}$ in spacetime, one may add an operator proportional to $i\phi^{*}\left(
v^{\mu}\partial_{\mu}\right)^{3}\phi$ to the Lagrange density for a scalar
particle. If $v^{\mu}$ has a time
component only, this will add a term proportional to $E^{3}$
to the usual relativistic energy-momentum relation $E^{2}=\vec{p}\,^{2}+m^{2}$.
Since the statement that $v^{\mu}$ is purely timelike is not Lorentz
invariant, this condition must be taken to hold is a particular preferred
frame, which is typically assumed to be
the rest frame of the cosmic microwave background. In this framework, the
electromagnetic field is incorporated through the usual minimal coupling
procedure. In the presence of this kind of Lorentz violation, the motion of a
charged particle in a constant magnetic field is modified, but the projection
of the trajectory onto
the plane perpendicular to $\vec{B}$ remains circular, and the particle's speed
remains constant. The radiation in the far field can be determined (including
information about polarization) and circumstances that could enhance observable
effects have been identified~\cite{ref-motemayor1,ref-motemayor2}.

Stringent bounds on Lorentz violations with modified dispersion
relations have been obtained from data from the Crab nebula~\cite{ref-jacobson1,
ref-jacobson2,ref-ellis}. These modifications
can lead to maximum particle velocities that are less than the speed of light,
but the Crab nebula shows evidence of synchrotron emission from electrons with
Lorentz factors of $\gamma=\left(1-\vec{v}\,^{2}\right)^{-1/2}\sim3\times10^{9}$.
For electrons with the conventional dispersion relation, this corresponds to
energies of 1500 TeV. The
existence of electrons with velocities this large can be used to constrain the
dispersion relation
models. If the coefficient of the Lorentz-violating operator in the Lagrangian
has a particular sign, the data show that it must be at least
seven orders of magnitude smaller than ${\cal O}(E/M_{P})$ Planck-level
suppression. This method, of placing bounds on the size of the Lorentz violation
based on the inferred velocities of astrophysical electrons, is the same one we shall
use here. However, we shall apply the arguments to more important superficially
renormalizable operators.

If spacetime is noncommutative, this will result in Lorentz noninvariant
physics~\cite{ref-carroll3}. Synchrotron processes have also been examined within
this framework. Because of the noncommutivity of the coordinates, there
automatically
are modifications to all sectors of the theory. The electron sector, the free
photon sector, and the minimal coupling between the two are all affected.
The discussion
in~\cite{ref-castorina} covers the particular case in which a magnetic
field and the Lorentz-violating noncommutativity parameter are aligned, so that
the orbits of charged particles in the plane perpendicular to $\vec{B}$ are
still given by circles. It is possible to work out the far fields within this
model at leading order in the noncommutativity. However, there are a number of
potential difficulties associated with the interpretation; these include
acausality and problems with quantization.

In section~\ref{sec-SME}, we shall introduce the SME terms that are likely to make
the largest contributions to observable high-energy astrophysical processes.
We shall explain why
all other terms should probably contribute negligibly in comparison,
but we shall also discuss how the same experimental data we
shall utilize could nonetheless be used to constrain other Lorentz-violating models.
Our bounds on Lorentz violation are related to the high-energy behavior of
electrons' velocities. The Lorentz-violating modifications of the velocity are
worked out in section~\ref{sec-vel}. There are two important effects. The maximum
electron speed might be less than one; or it might be greater than one, so that there
is a maximum energy for subluminal electrons. In section~\ref{sec-proc}, we discuss
the necessary details of the astrophysical synchrotron and inverse Compton
processes, and in section~\ref{sec-analysis}, we describe how parameters such as
the maximum electron speed can be inferred from observed spectral data. Then in
section~\ref{sec-data}, we look at the data that is actually available and use it
to place bounds on the Lorentz-violating coefficients. Finally, we discuss our
conclusions and the prospects for further improvement in section~\ref{sec-concl}.

\section{Lorentz-Violating QED}

\label{sec-SME}

\subsection{The $c$ Term}

Both modified dispersion relation and noncommutative spacetime theories
involve Lo\-rentz-violating operators that are nonrenormalizable by power counting.
On the other hand, there is a unique spin-independent, superficially renormalizable
SME coupling that is consistent with the
gauge invariance of the standard model and which grows in relative
importance at high
energies. This is a CPT-even two-index tensor $c^{\nu\mu}$, and this is the term
which we will seek to constrain.

The Lagrange density we shall consider is
\begin{equation}
{\cal L}=-\frac{1}{4}F^{\mu\nu}F_{\mu\nu}+\bar{\psi}[\Gamma^{\mu}(i\partial_{\mu}
-eA_{\mu})-M]\psi,
\end{equation}
where
$\Gamma^{\mu}=\gamma^{\mu}+c^{\nu\mu}\gamma_{\nu}$ and $M=m$, so that
\begin{equation}
\label{eq-L}
{\cal L}=-\frac{1}{4}F^{\mu\nu}F_{\mu\nu}+
\bar{\psi}[(\gamma^{\mu}+c^{\nu\mu}\gamma_{\nu})
(i\partial_{\mu}-eA_{\mu})-m]\psi.
\end{equation}
Here, $\psi$ is the electron field, and 
$c$ contains nine parameters that contribute to Lorentz-violating physics at leading
order. The expression
(\ref{eq-L}) is not the full Lagrange density for the electron and photon
sectors of the minimal SME. More generally, we could have
\begin{equation}
\Gamma^{\mu}=\gamma^{\mu}+c^{\nu\mu}\gamma_{\nu}-d^{\nu\mu}\gamma_{\nu}
\gamma_{5}+e^{\mu}+if^{\mu}\gamma_{5}+\frac{1}{2}g^{\lambda\nu\mu}
\sigma_{\lambda\nu}
\end{equation}
and
\begin{equation}
M=m+\!\not\!a-\!\not\!b\gamma_{5}+\frac{1}{2}H^{\mu\nu}\sigma_{\mu\nu}+im_{5}
\gamma_{5},
\end{equation}
as well as terms $-\frac{1}{4}k_{F}^{\mu\nu\rho\sigma}
F_{\mu\nu}F_{\rho\sigma}$ and $\frac{1}{2}k_{AF}^{\mu}\epsilon_{\mu\nu\rho\sigma}
F^{\nu\rho}A^{\sigma}$ in the electromagnetic sector. However, 
the $c$
coefficients are the only sources of Lorentz violation that we shall need to
consider. As we shall
discuss shortly, all other Lorentz-violating terms can either be absorbed into
$c$ or will have their contributions suppressed.

The trace $c^{\mu}\,_{\mu}$ only affects the overall normalization of the
electron
field, and the antisymmetric part of $c$ has no effects at first order, because at
that order, it
is equivalent to a change in the representation of the Dirac matrices. 

We shall choose a $c$ that is not symmetric, but rather one with 
$c^{\nu0}=0$. This can be accomplished using a field redefinition. We do this because
it will simplify our calculations. The utility of this choice comes from the fact
that, despite the Lorentz violation, the electromagnetic field is coupled
conventionally to the velocity via a $v^{\mu}A_{\mu}$ term. Because $c^{\nu0}=0$,
the electrostatic potential $\Phi=A^{0}$ is
coupled, as usual, to the charge density $e\psi^{\dag}\psi$, and the
vector potential $\vec{A}$ couples to $e\psi^{\dag}\dot{\vec{x}}\psi$. (Although
the operator $e\psi^{\dag}\dot{\vec{x}}\psi$ technically contains
{\em Zitterbewegung}, it will be perfectly valid to ignore this effect.)
So all the usual results for the electromagnetic field of a moving pointlike charge
continue to hold, once the charge's motion is prescribed. Determining this motion
is also relatively simple, because the equation of motion for the particle is the
unmodified Lorentz force law.

Moreover,
the canonical quantization of the fermion field requires some care when certain
$c$ coefficients are nonvanishing. If $c^{\nu0}$ were nonzero, then
${\cal L}$ would contain nonstandard time derivative terms.
In that case, a matrix transformation $\psi'=R\psi$ would be required,
to ensure that $\Gamma^{0}=\gamma^{0}$. An explicit power-series expression for
the required $R$ is given in~\cite{ref-lehnert3}.
For simplicity, we shall assume that any such necessary
transformation has already been performed and $c^{\nu0}=0$.
However, this will require us to consider the canonical quantization in a single
frame only. We may not boost the theory into another frame, because doing so
would reintroduce the problematic time derivatives. The
precise frame in which we shall take $c^{\nu0}=0$ is the standard sun-centered
celestial equatorial coordinate frame that is used in the study of Lorentz
violation~\cite{ref-bluhm4}.

The $c$ coefficients for protons are generally more tightly constrained than those
for electrons. The proton's $c$ values can be measured by precision atomic clock
comparison experiments. Unfortunately, the hyperfine transitions used in these
measurements are not sensitive to $c$-type Lorentz violation in the electron
sector. However, other atomic transitions at much higher energies may be sensitive to
the electron $c$ terms; future laboratory experiments could use measurements of
those transitions to constrain many of the same terms that are being discussed here.

\subsection{Terms to Be Neglected}

There are
other superficially renormalizable couplings contained in the minimal SME,
but as already noted, the $c$ couplings are most natural in this context, and
they should
make the largest contributions to the effects we plan to study.
Nonrenormalizable operators---which are outside the minimal SME---we naturally
expect to be suppressed relative to
renormalizable ones by powers of a large momentum scale. When
considering synchrotron and IC radiation, one is primarily interested
in particles
with very high energies. Lorentz-violating coefficients that modify the kinetic
part of the Lagrangian will grow in relative
importance at high energies, as the components of the momentum become
large, so it is natural
to consider only such kinetic modifications; thus we may neglect $a$, $b$, and $H$,
whose effects do not increase with energy. 
There remain in $\Gamma^{\mu}$ only two sets of Lorentz-violating
that are consistent with the actual standard model's chiral
gauge couplings---the $c^{\nu\mu}$ terms and also
the set of $d^{\nu\mu}$ terms, which have the same form as the $c$ interactions,
except for the addition of a $\gamma_{5}$.

However, we do not consider the $d$ interactions here, because their effects are
expected to be small. Contributions from $d$ should average out,
because there should be no net polarization of the electrons in high-energy
sources. In fact, for
an electron undergoing circular cyclotron motion, with the spin oriented in
the plane of the orbit, the spin rotates by
$2\pi\gamma\frac{g-2}{2}$ radians with each orbital revolution. For
$\gamma\gg\alpha^{-1}$, the spin will rotate many times during one orbital
period, and any effects proportional to the helicity will be diminished by the
resultant averaging. Moreover,
mixing between the standard kinetic term and the $c$ term
(which have the
same basic Dirac structure) also causes the
effects of $c$ also grow in importance relative to those of
$d$ at high energies~\cite{ref-kost3}. If the scale of the dimensionless
coefficients $c$ and $d$ is ${\cal O}(m/M_{P})$, where $M_{P}$ is some large
momentum scale, then the effects of $c$ grow to be large at momenta $|\vec{p}\,|
\sim\sqrt{mM_{P}}$, while those of $d$ grow more slowly, becoming large only when
$|\vec{p}\,|\sim M_{P}$. [If there is physical Lorentz violation, then $M_{P}$ may
represent the Planck scale. However, for the present purposes, we may take $M_{P}$
to be effectively
defined by the size of $c$; $M_{P}$ is whatever scale is needed so that
$c$ will be ${\cal O}(m/M_{P})$.]

Modifications of the kinetic Lagrangian that are not invariant under the standard
model's $SU(2)_{L}$ gauge symmetry can also exist; these are $e$, $f$, and $g$.
However, these can only appear
as part of electroweak symmetry breaking, as vacuum expectation values of
higher dimensional (i.e., nonrenormalizable) operators. These operators should
therefore be further
suppressed, and we shall also neglect them.

Finally, we may also neglect any Lorentz
violation in the photon sector. Modifications of the free electromagnetic
Lagrangian will generally change the speed of photon propagation. Most possible
Lorentz-violating terms in the free
electromagnetic sector (all the components of $k_{AF}$ and half those of $k_{F}$)
give rise to photon birefringence, because photons with right-
and left-handed circular polarizations travel at different speeds. This
birefringence has been
searched for and not seen. The limits on the
relevant forms of Lorentz violation are very
strong, and we may safely neglect them. The purely electromagnetic
terms that do not cause birefringence can be accounted for by adding 
\begin{equation}
{\cal L}_{F}=-\frac{1}{4}\left(k_{F}\right)^{\alpha}\, _{\mu\alpha\nu}
\left(F^{\rho\mu}F_{\rho}\,^{\nu}+F^{\mu\rho}F^{\nu}\, _{\rho}\right).
\end{equation}
to ${\cal L}$. However, a coordinate transformation $x^{\mu}\rightarrow x^{\mu}
-\frac{1}{2}\left(k_{F}\right)^{\alpha\mu}\,_{\alpha\nu}x^{\nu}$ will eliminate
all the Lorentz violation from the photon sector at leading
order~\cite{ref-kost16,ref-kost17}. This transformation shifts
the Lorentz-violating physics into the charged matter sector, where it
manifests itself exactly as a $c^{\nu\mu}$ term. So we see that consideration of $c$
captures all the possible sources of Lorentz violation in a
synchrotron process that are not significantly further suppressed.
However, it is important to note that the transformation that eliminates $k_{F}$ is
frame-dependent, and the new coordinates need not even be rectangular relative to
the original ones; so by choosing to consider only this form of Lorentz violation,
we are restricting ourselves to working in a very particular and special
coordinate system.

\subsection{Alternative Formulations}

\subsubsection{Alternative Form for $c$}

In many applications, it is more convenient to consider a $c^{\nu\mu}$ which is
both traceless and symmetric. We shall refer to the equivalent $c$ tensor with these
properties as $c_{TS}$, and all the bounds given here can be easily translated into
bounds on the coefficients of $c_{TS}$.

The trace $c^{\mu}\,_{\mu}$ can actually
be eliminated from the theory to all orders by a field redefinition. $g^{\nu\mu}+
c^{\nu\mu}$ is a bilinear form that connects $\gamma_{\nu}$ and $p_{\mu}$ in the
action. However, its trace can be fixed to
four by inserting a new set of fermion fields
$\psi'=\sqrt{1+\frac{1}{4}c^{\mu}\,_{\mu}}\psi$. (This field redefinition must be
accompanied by a change in the value of the mass parameter $m$; however, this change
is unimportant for the ultrarelativistic phenomena considered here.)
At leading order, the redefinition changes
$c^{\nu\mu}$ to $c^{\nu\mu}-\frac{1}{4}c^{\alpha}\,_{\alpha}g^{\nu\mu}$. Symmetrizing
this expression is trivial to accomplish, so the
final traceless symmetric $c_{TS}$
equivalent (at leading order) to our $c$ is
\begin{equation}
c_{TS}^{\nu\mu}=\frac{1}{2}\left(c^{\nu\mu}+c^{\mu\nu}\right)
-\frac{1}{4}c^{\alpha}\,_{\alpha}g^{\nu\mu}.
\end{equation}

In fact, clock comparison experiments, which set the best bounds on the $c$
coefficients for protons and neutrons, have only bounded such combinations as
$c_{(XY)}$ [where the parentheses denote the symmetrized expression
$c_{(XY)}=c_{XY}+c_{YX}$], $c_{0X}$, $c_{XX}-c_{YY}$, and
$c_{Q}=c_{XX}+c_{YY}-2c_{ZZ}$.
(The coordinate system in which $X$, $Y$, and $Z$ are defined is explained in
section~\ref{sec-data}.)
None of these combinations of coefficients
is actually sensitive to whether or not $c$ is traceless. These experiments only
constrain eight of the nine physical $c$ coefficients per species.
Clock comparison bounds on
the remaining nucleon coefficients---the $(c_{TS})_{00}$ [or equivalently, the
$(c_{TS})_{jj}$]---have not been calculated, because they would
be suppressed by two powers of the Earth's
revolution speed, $v_{\oplus}\sim 10^{-4}$.
Moreover, the bounds on the nucleons' $c_{0j}$ terms are already worse than the
bounds on the other
coefficients by one power of $v_{\oplus}$. In general, for each time index on
an element of $c$, the observable effects (which come from violations of boost
invariance) are suppressed by one power of the velocities involved. For laboratory
experiments, where the largest speed available is $v_{\oplus}$, this is a major
impediment. However, when the bounds are based on observations of
ultrarelativistic electrons, this fact presents no problem at all, because the
particles' speeds are all very close to one. Therefore, we can find bounds on
all the physically meaningful electron $c$ coefficients at the same level of
accuracy; any differences in the bounds are due solely to the differing the quality
of the data available for sources in various directions.

\subsubsection{Other Theories}

We shall be bounding the $c$ coefficients for electrons by looking at the structure
of the theory near the electrons' maximum velocities. Obviously, without any Lorentz
violation, this maximum velocity is one. However, a $c$ term will generally result
in a change to the maximum speed. If $c_{jk}\propto\delta_{jk}$, the electrons'
limiting speed is the same in all directions; the free electron theory looks just
like ordinary special relativity, but with a different value for the speed of light.
As we will see, a $c_{0j}$ term results is a similar distortion of the
energy-momentum relation, with the added wrinkle that the maximum speed is
direction-dependent. This
angular dependence is dipolar; for electrons travelling in a
direction $\hat{e}$, the maximum speed is $1-c_{0j}\hat{e}_{j}$. An electron
moving the opposite direction will have a different maximum speed, and in the plane
normal to the three-vector defined by the $c_{0j}$, the top speed is one. The
traceless part of $c_{jk}$ has a similar effect, except that the deformation of
the maximum speed has a quadrupole pattern. (We shall derive all these facts in
section~\ref{sec-vel}.)

So $c$ leads to three kinds of deviations in the maximum election speed, which
have three of the most obvious possible forms. There are other
possibilities too, however. Direction-dependent changes in the speed of light
with octopole or higher multipole characteristics are not possible with just a $c$
term, but they might occur in more general theories. Modifications
of Lorentz invariance which are only important above some large scale $M$, such
as would result from
a deformed energy-momentum relation $E^{2}=m^{2}+\vec{p}\,^{2}+|\vec{p}\, |^{3}
/M$ are also possible. However, regardless of the structure of a specific theory,
if the electrons' maximum speed is different from one, bounds of the sort
discussed here will be available, based on the very same observations.
As already mentioned, for the
theory with the $|\vec{p}\, |^{3}/M$ modification, some of these bounds have already
been worked out~\cite{ref-jacobson1}. We shall not pursue the analyses of such
alternative theories any further, since the effective field theory operators that
give rise to these kinds of Lorentz violations
are generally exotic and nonrenormalizable, making them less important than $c$.
However, we should keep in mind that bounds similar to the ones here are possible
in these nonrenormalizable theories as well.

\section{Electron Velocities}

\label{sec-vel}

\subsection{Derivation of $\vec{v}$}

Because the free electromagnetic sector and the coupling to charged matter in
(\ref{eq-L}) are
completely conventional, standard effects in electrodynamics may be used as sensitive
probes of the Lorentz-violating electron sector. In particular, we may place strong
bounds on $c$ by looking at the relationships between energy, momentum, and velocity
in the $c$-modified theory. At ultrarelativistic energies, the bounds on $c$ that
can be derived from observing the emissions of an electron with Lorentz factor
$\gamma$ go as $\gamma^{-2}$. This strong dependence on $\gamma$ is a consequence of
the rapid growth in the importance of $c$ with energy. If $c$ is
${\cal O}(m/M_{P})$, its effects will become important at scale
$E\sim\sqrt{mM_{P}}$, and the Lorentz factor at this scale is
$\gamma\sim\sqrt{M_{P}/m}$. So if no effects of Lorentz violation are observed up to
some Lorentz factor $\gamma$, this constrains $c$ to be smaller than
${\cal O}(\gamma^{-2})$. (On the other hand, any bounds
on the $d$, $e$, $f$, and $g$ coefficients would scale as $\gamma^{-1}$, at best.)

We must understand the effects of Lorentz violation in the free electron sector in
order to place constraints on $c$. So we shall
for the moment neglect the electromagnetic coupling
and just look at how the free
electrons behave. In particular, we shall look at the structure
of the electron velocity. This means doing single-particle relativistic quantum
mechanics, starting from a modified Dirac equation~\cite{ref-altschul4}. We
shall see that the conventional relations between energy, momentum, and
velocity---$E=\gamma m$ and $\vec{\pi}=\gamma m\vec{v}$---no longer hold because of
the Lorentz violation.

It turns out however, that even in the presence of any of $c$, the velocity may be
found exactly. With the Lorentz violation, the single-particle Hamiltonian becomes
\begin{equation}
H=\alpha_{j}\pi_{j}-c_{lj}\alpha_{l}\pi_{j}-c_{0j}\pi_{j}+\beta m.
\end{equation}
As usual, the Dirac matrices are $\alpha_{j}=\gamma^{0}\gamma^{j}$ and
$\beta=\gamma^{0}$. $\vec{\pi}$ is the mechanical momentum, which coincides with
the canonical momentum in the free case; however when we reintroduce the
electromagnetic coupling, the velocity will depend on $\vec{\pi}=\vec{p}-e\vec{A}$,
rather than the gauge-noninvariant canonical momentum $\vec{p}$.
The Hamiltonian $H$ generates
the Heisenberg equation of motion for the position operator $x_{k}$,
\begin{equation}
\dot{x}_{k}=i[H,x_{k}]=\alpha_{k}-c_{lk}\alpha_{l}-c_{0k};
\end{equation}
so in the presence of the
Lorentz violation, $\vec{\alpha}$ is no longer the velocity operator,
although the velocity is still an affine function of the $\alpha_{j}$.

To find the velocity, we must solve the Heisenberg equation of motion for
$\alpha_{k}$. This equation is
\begin{equation}
\dot{\alpha}_{k}=i\left[-2\alpha_{k}(H+c_{0j}\pi_{j})+2\pi_{k}-2c_{kj}\pi_{j}\right],
\end{equation}
and it has an exact solution analogous to the Lorentz-invariant one---
\begin{equation}
\label{eq-alpha}
\alpha_{k}(t)=\left(\pi_{k}-c_{kj}\pi_{j}\right)(H+c_{0j}\pi_{j})^{-1}+
\left[\alpha_{k}(0)-\left(\pi_{k}-c_{kj}\pi_{j}\right)(H+c_{0j}\pi_{j})^{-1}\right]
e^{-2i(H+c_{0j}\pi_{j})t}.
\end{equation}
$H$ and $\vec{\pi}$ are constants of the motion. The second term on the right-hand
side of (\ref{eq-alpha}) is
matrix-valued and oscillatory. It is a {\em Zitterbewegung} term,
completely analogous to the one found in the Lorentz-invariant case.
The {\em Zitterbewegung} arises from interference between positive- and
negative-frequency plane waves, and it is solely responsible for the fact that the
components of the velocity do not commute with one-another. The {\em Zitterbewegung}
is purely quantum-mechanical in origin, and it may be neglected. The
{\em Zitterbewegung} motion can be eliminated entirely if the electron wave packet
contains no negative-frequency components, which is possible if the particle is
spread out spatially, with a position uncertainty larger than the Compton wavelength
$1/m$. Moreover, to the extent that we want to consider the radiation from electrons
with well-defined velocities, we must drop the {\em Zitterbewegung}, because it
prevents us from resolving more than one component of the velocity at a time.

The {\em Zitterbewegung}-free contribution to the velocity is therefore
\begin{equation}
\label{eq-v}
v_{k}=\frac{1}{E+c_{0j}\pi_{j}}\left(\pi_{k}-c_{kj}\pi_{j}-c_{jk}\pi_{j}+c_{jk}
c_{jl}\pi_{l}\right)-c_{0k}.
\end{equation}
The Hamiltonian has been replaced by its eigenvalue $E$. $E$ is the energy
corresponding to the momentum $\vec{\pi}$, $E=\sqrt{m^{2}+\left(\pi_{k}
-c_{kj}\pi_{j}\right)\left(\pi_{k}-c_{kl}\pi_{l}\right)}-c_{0j}\pi_{j}$. The group
velocity
derived from $E$ is the same as the {\em Zitterbewegung}-free $\vec{v}$; however,
the algebraic method of deriving the velocity is more general.

To first order, only the symmetric part of $c_{kj}$ contributes to $\vec{v}$, as
expected. The antisymmetric part corresponds at this order merely to a change
in the representation of the $\gamma_{j}$ Dirac matrices; such a change can
have no physical consequences. [The fact that (\ref{eq-v}) depends asymmetrically
on $c_{0j}$ and not $c_{j0}$ is a consequence of the fact that our calculational
methods have already made use of the fact that $c^{\nu0}=0$.]

We can now see why the effects of $c$ become large at the scale $\sqrt{mM_{P}}$.
According to (\ref{eq-v}), the velocity might become superluminal when
$|\vec{\pi}|/E\approx 1-|c|$, where $|c|$ is a characteristic size for the
Lorentz-violating
coefficients. This gives us an estimate of the maximum value of $\gamma$ that
can be achieved before new physics must come into play if some form of causality
is to be preserved. For ultrarelativistic particles,
$\gamma\approx\left[2\left(1-|\vec{\pi}|/E\right)\right]^{-1/2}$, and this
diverges at an energy scale $E_{\max}\sim m/\sqrt{|c|}\sim\sqrt{mM_{P}}$. Above this
scale, 
the description of the Lorentz violation through
an effective field theory containing only $c^{\nu\mu}$ terms will generally
break down, and higher dimension operators should become important.

\subsection{Maximum Velocity}

We can exploit the Lorentz violation in the relationship between momentum and
velocity to place bounds on $c$. There are two crucial and complementary effects.
The first effect is that the maximum electron speed in a given direction
$\hat{e}$ is generally different from one. The value of this new maximum velocity
can be easily calculated to first order in $c$. We simply expand $v_{j}\hat{e}_{j}$
to ${\cal O}(c)$, finding
\begin{equation}
\label{eq-vdote}
v_{j}\hat{e}_{j}=\frac{1}{\sqrt{m^{2}+\pi_{j}\pi_{j}}}\left(\pi_{j}\hat{e}_{j}+c_{jk}
\frac{\pi_{j}\pi_{k}}{m^{2}+\pi_{l}\pi_{l}}\pi_{l}\hat{e}_{l}-c_{jk}\pi_{j}
\hat{e}_{k}-c_{jk}\pi_{k}\hat{e}_{j}\right)-c_{0j}\hat{e}_{j}.
\end{equation}
At large momenta, we may neglect the mass $m$. Then in terms of the unit vector
$\hat{\pi}$ in the momentum direction, $v_{j}\hat{e}_{j}$ becomes
\begin{equation}
v_{j}\hat{e}_{j}=\hat{\pi}_{j}\hat{e}_{j}+c_{jk}\hat{\pi}_{j}\hat{\pi}_{k}
\hat{\pi}_{l}\hat{e}_{l}-c_{jk}\hat{\pi}_{j}\hat{e}_{k}-c_{jk}\hat{\pi}_{k}
\hat{e}_{j}-c_{0j}\hat{e}_{j}.
\end{equation}
The speed will be maximum when the momentum direction is $\hat{\pi}=\hat{e}+{\cal O}
(c)$. In the terms that already contain $c$, the $c$-dependent corrections to this
direction may be neglected; we may therefore replace $\hat{\pi}$ with $\hat{e}$ in
the ${\cal O}(c)$ terms and then maximize the resulting expression, which is just
$\hat{\pi}_{j}\hat{e}_{j}-c_{jk}\hat{e}_{j}\hat{e}_{k}-c_{0j}\hat{e_{j}}$.
This is clearly still maximized by setting $\hat{\pi}=\hat{e}$, as in the
Lorentz-invariant case. So the limiting value of $v_{j}\hat{e}_{j}$ is
\begin{equation}
\label{eq-vmax}
\left(v_{j}\hat{e}_{j}\right)_{\max}=1-c_{jk}\hat{e}_{j}\hat{e}_{k}-c_{0j}
\hat{e}_{j}.
\end{equation}
If this is less than one, it can have readily observable consequences.

\subsection{Maximum Subluminal Energy}

The complementary effect is that there may be a maximum energy available to
electrons with subluminal velocities. This will also generally depend on the
direction of a particle's motion. To determine this energy, we again use
(\ref{eq-vdote}), setting $v_{j}\hat{e}_{j}=1$.
However, in this case the mass cannot be neglected; instead, we
expand to leading order in $m^{2}/\pi_{j}\pi_{j}$. This gives
\begin{equation}
1+\frac{m^{2}}{2\pi_{j}\pi_{j}}=\hat{\pi}_{j}\hat{e}_{j}
+c_{jk}\hat{\pi}_{j}\hat{\pi}_{k}\hat{\pi}_{l}\hat{e}_{l}\left(1-\frac{m^2}
{\pi_{j}\pi_{j}}\right)-c_{jk}\hat{\pi}_{j}\hat{e}_{k}-c_{jk}\hat{\pi}_{k}\hat{e}_{j}
-c_{0j}\hat{e}_{j}.
\end{equation}
We know from the previous
calculation that the velocity in the $\hat{e}$-direction will reach the speed of
light most easily (that is, with the smallest momentum) when $\vec{\pi}$ and
$\hat{e}$ are aligned. Any corrections to this alignment are at least second order
in $c$. So we may replace $\hat{\pi}$ with $\hat{e}$ everywhere.
Doing this, we have
\begin{equation}
\label{eq-solvingE}
\frac{m^2}{2\pi_{j}\pi_{j}}\left(1+2c_{jk}\hat{e}_{j}\hat{e}_{k}\right)=-c_{jk}
\hat{e}_{j}\hat{e}_{k}-c_{0j}\hat{e}_{j}.
\end{equation}
Up to corrections of ${\cal O}(c)$ or ${\cal O}(m^{2}/\pi_{j}\pi_{j})$,
$\pi_{j}\pi_{j}/m^{2}$ is simply $E^{2}/m^{2}$. The corrections just mentioned, as
well as the
$c$-dependent terms on the left-hand side of (\ref{eq-solvingE}), may be neglected.
They are small corrections to the expression
\begin{equation}
\label{eq-Emax}
\frac{E}{m}=\frac{1}{\sqrt{-2c_{jk}\hat{e}_{j}\hat{e}_{k}-2c_{0j}\hat{e}_{j}}}.
\end{equation}
The maximum subluminal energy $E$ is proportional to the inverse square root of $c$,
which is not surprising, since for vanishing $c$, $E$ must be infinite. As is
obvious from (\ref{eq-Emax}), such a maximum value for $E$ need not always exist.
In fact, according to (\ref{eq-vmax}), this maximal subluminal energy does not
exist precisely when the maximum speed in the relevant direction is less than or
equal to one, and this is exactly what one would expect.

\subsection{Relationship to Bounds on $c$}

The two conditions we have found, that $v_{j}\hat{e}_{j}$ cannot exceed
$1-c_{jk}\hat{e}_{j}\hat{e}_{k}-c_{0j}\hat{e}_{j}$ and that $E/m$ cannot
exceed $\left(-2c_{jk}\hat{e}_{j}\hat{e}_{k}-2c_{0j}\hat{e}_{j}\right)^{-1/2}$
without an
electron becoming superluminal, look very different, although they do both
depend on $c$ through the expression
$c_{jk}\hat{e}_{j}\hat{e}_{k}+c_{0j}\hat{e}_{j}$. However, we these two conditions
can actually be cast in very similar forms, when they are related to experimental
observations. If we separately observe the existence of electrons with Lorentz
factors up to some value $\gamma_{\max}=\left(1-v_{\max}^{2}\right)^{-1/2}$ moving
in the $\hat{e}$-direction and subluminal electrons with energies up to $E_{\max}$
traveling in the same direction, then this restricts
$c_{jk}\hat{e}_{j}\hat{e}_{k}+c_{0j}\hat{e}_{j}$ to lie in the range
\begin{equation}
\label{eq-2sidebound}
-\frac{1}{2\left(E_{\max}/m\right)^{2}}<
c_{jk}\hat{e}_{j}\hat{e}_{k}+c_{0j}\hat{e}_{j}<\frac{1}{2\gamma_{\max}^{2}}.
\end{equation}

Naturally, in the absence of Lorentz violation, $\gamma=E/m$. By measuring
$\gamma_{\max}$ and $E_{\max}$, we are constraining the electrons' energy-momentum
relation. The speed of light is fixed to be one by the conventional electromagnetic
sector. This provides a basis for comparison when we search for Lorentz violations
in other sectors. If Lorentz
symmetry is exact, the maximum possible $\gamma$ and $E$ are both infinite.
Yet this is not generally the case with Lorentz violation, and the Lorentz-violating
$c$ terms distort the behaviors of the Lorentz factor and the energy differently.
So it is not surprising that the bounds derived from $\gamma_{\max}$ and
$E_{\max}$ are different, and it is quite convenient that they are actually
so complementary.

\section{High-Energy Radiation Processes}

\label{sec-proc}

\subsection{Synchrotron Process}

The classical synchrotron process involves electrons revolving helically around
lines of magnetic flux. These accelerated particles emit radiation over a broad
spectrum of frequencies, up to a characteristic cutoff. We shall review the crucial
features of this phenomenon, emphasizing those characteristics that will be
important for our study of Lorentz violation. An excellent source that discusses
the importance a various effects in determining astrophysical synchrotron spectra
is~\cite{ref-pacholczyk}, although the basic material can be found in many
treatments.

We have previously presented a much more detailed account (to all orders in $c$),
of the synchrotron process in the presence of Lorentz violation~\cite{ref-altschul5}.
However, most of the detailed results arising from this treatment
turn out to be unimportant for our
attempt to set bounds on $c$. The predominant effects are the ones discussed in
section~\ref{sec-vel}, which were derived only from the electrons' energy-momentum
relation. There are some other interesting
qualitative changes in the behavior of the system that we shall mention; however, we
shall not treat them in great detail.

Thus far, we have used rationalized units for the electromagnetic field and charge.
However, for explicit calculations involving electrodynamics with moving sources, it
is easier and more conventional to use Gaussian units. We shall use these
unrationalized units
in sections~\ref{sec-proc} and~\ref{sec-analysis}, so that the formulas will look
more familiar.

\subsubsection{Synchrotron Motion}

The basic phenomenon of synchrotron motion is well known. Charged particles
(in this case electrons) moving a high speeds are accelerated perpendicular to
the direction of their motion by a basically homogeneous magnetic field. Their
trajectories curve around the magnetic field lines. This confines the electrons
in the plane perpendicular to $\vec{B}$, although they still move freely along the
direction of the field. This picture changes only slightly when a Lorentz-violating
$c$ term is added.

At the energies we are interested in, all quantum effects can be neglected. This
applies to both the electrons' motions and, as we shall see, to the radiation
emissions. Quantum effects could enter through Landau level quantization or
spin-dependent effects, or through the {\em Zitterbewegung}. The spin-orbit and
Landau level effects make the spectrum discrete, but the difference between adjacent
energy levels is miniscule compared to the energies involved. Since the quantum
numbers are very large, the classical treatment is an excellent approximation.
Similarly, although the velocity that the three-vector potential couples to is,
in principle, the 
the velocity $e\psi^{\dag}\dot{\vec{x}}\psi$ including
the {\em Zitterbewegung}, we know empirically that {\em Zitterbewegung} is
unimportant in the emission of astrophysical synchrotron radiation. The changes to
the {\em Zitterbewegung} due to the Lorentz violation are minor, and result in no
qualitative differences that would suddenly make this quantum
interference effect important. So 
we may treat the electromagnetic field as if it were coupled simply to the
group velocity $\vec{v}$.

The classical result is that an electron's motion perpendicular to $\vec{B}$ is
circular, with
angular revolution frequency $\omega_{B}=\left|e\vec{B}\,\right|/\gamma m$. The
radius of
gyration is $\rho=\left|\vec{p}\times\vec{B}\,\right|/|e|\vec{B}^{2}\approx
E\sin\theta/\left|e\vec{B}\,\right|$, where $\theta$ is the angle between the
magnetic field and
the direction of the electron's motion. The frequency is only slightly modified
by the Lorentz-violating $c$. Depending on which formula is used, the radius
$\rho$ may be more significantly modified; however, this effect turns out not to
be important. Lorentz violation will generally also shift the periodic part of the
electron's motion out of the plane normal to $\vec{B}$; the circular path in the
plane becomes an elliptical one tilted out of the plane, and the electron's
velocity depends on time, varying with an
angular frequency $2\omega_{B}$. The deviations
in the shape and orientation of the orbit are ${\cal O}(c)$ and can be neglected.
Moreover, although changes to the velocity of the emitting electrons are exactly
what we want to measure, we cannot observe the
periodic time variations in $\vec{v}$ directly, because of the way the
synchrotron radiation is emitted.

\subsubsection{Synchrotron Radiation}

As an electron revolves in the magnetic field, it emits radiation, which, because
of the particle's
ultrarelativistic velocity, is beamed into a narrow pencil of angles
around the instantaneous direction of the velocity. The characteristic angular
spread of the emission is ${\cal O}(\gamma^{-1})$, and this width is generally
neglected; instead, we treat all the radiation as if it were emitted precisely
along the tangent vector to the trajectory. The intensity of the
radiation caused by any accelerations parallel to the velocity is smaller than the
synchrotron radiation by a factor of ${\cal O}(\gamma^{-2})$, provided the
accelerations are comparable. Moreover, this radiation is beamed into the same
narrow pencil of angles as the synchrotron emission, so it has no angular properties
to distinguish it, and it has no meaningful effect on the observable spectrum.

The frequency spectrum is discrete, all the power being radiated in harmonics of
the fundamental frequency $\nu_{0}=\frac{\omega_{B}}{2\pi}\csc^{2}\theta$.
The emitted power is spread over all harmonics less
than the critical frequency
\begin{equation}
\label{eq-nuc}
\nu_{c}=\frac{3}{4\pi}\gamma^{3}\omega_{B}\sin\theta
=\frac{3}{4\pi m}\gamma^{2}\left|e\vec{B}\,\right|\sin\theta.
\end{equation}
Most of the power is emitted close to this frequency, and above
$\nu_{c}$, the radiated power falls off very rapidly. The rate of energy loss of
an electron, found by summing the emission over all frequencies, is
\begin{equation}
\frac{dE}{dt}=-\frac{2e^{4}}{3m^{2}}\gamma^{2}\vec{B}^{2}\sin^{2}\theta.
\end{equation}

The $\nu_{c}$ in (\ref{eq-nuc}) represents
the critical frequency for the emissions of a single
electron. If a source contains significant numbers of electrons with velocities
up to some maximum Lorentz
factor $\gamma_{\max}$, then the observed cutoff in the spectrum
will be at the cutoff frequency for the most energetic electrons---$\nu_{c}=
\frac{3}{4\pi m}\gamma_{\max}^{2}\left|e\vec{B}\,\right|\sin\theta$. (Detailed
calculations for a truncated power law spectrum are given in~\cite{ref-pacholczyk}.)
This result for $\nu_{c}$ is what will allow us to infer $\gamma_{\max}$ from
observations of spectra.

Quantum effects are unimportant in the emission part of the synchrotron process,
just as they are in determining the electrons' trajectories.
The leading order quantum corrections to the standard synchrotron
formulas are negligible if
$2\pi\nu_{c}\ll E$, or equivalently if
$\gamma_{\max}\ll m^{2}/\left|e\vec{B}\,\right|$~\cite{ref-schwinger2}. 
Inserting the electron mass and charge, this is $\gamma\ll \left(3\times 10^{13}
\right)/\left|\vec{B}\,\right|$ if the magnetic field is measured in Gauss.
Since typical magnetic field strengths in synchrotron sources are fractions of
mG, the range of $\gamma$ values over which no quantum modifications are
necessary is extremely large. In particular,
the classical treatment will apply up to well beyond the scale of any observed
$\gamma_{\max}$.

\subsection{Inverse Compton Process}

The highest-energy photons that are emitted by astrophysical sources arise in
IC processes. Low-energy photons (which often come from synchrotron emission)
scatter off ultrarelativistic electrons. An electron may transfer a substantial
fraction of its own energy to a photon during such a collision, resulting in
the emission of photons whose energies may range almost up to the scale of
the highest electron energies.

The details of IC scattering can be worked out by taking the Klein-Nishina formula
and the usual kinematics of
Compton scattering (starting from a frame in which the electron is at rest) and
transforming into a frame where the electron is moving with a speed $v\approx 1$.
The details of the cross section will be unimportant in this instance. What matters
is the transformation of the kinematics, which we shall now examine. This
subject is covered in more detail in~\cite{ref-pacholczyk}, although the emphasis
there is not on precisely the same limit.

Collisions of photons with ultrarelativistic electrons are practically all head-on
when viewed in the electron's frame. To see this, we note that the
Lorentz transformation law for $\cos\psi$, where $\psi$
is the angle between some given direction (which we shall take to be the direction
of the photon's motion) and the boost direction (which is the direction of the
electron's motion, since we are boosting the electron from rest into an
ultrarelativistic frame) is
\begin{equation}
\cos\psi_{i}^{e}=\frac{\cos\psi^{o}_{i}-v}{1-v\cos\psi^{o}_{i}},
\end{equation}
where the superscripts denote quantities taken in the rest frame of the observer
($o$) or of the electron
($e$), and the subscripts denote that these are initial values, applying prior to the
scattering.
As $v\rightarrow1$, $\cos\psi^{e}\rightarrow -1$, so
$\psi_{i}^{e}\approx\pi$. Overtaking collisions are extremely rare, occurring only in
a miniscule range of observer frame solid angles.

We are interested in collisions where the
fractional change in the photon energy is substantial. In such collisions, the
photon's scattering angle
$\theta^{e}$ in the electron's rest frame can never be small. When we boost into
the observer's frame, all vectors that are not aligned almost perfectly antiparallel
to the boost direction lie, after the boost, within a small pencil of angles about
the boost direction. The fact that $\theta^{e}$ is not small just
means that the outgoing
photon is not moving nearly antiparallel to the electron's initial direction;
therefore, it rebounds back with a very large observer frame scattering angle,
$\theta^{o}\approx\pi$. So the emitted photon is propagating in essentially the
same direction as the initial electron.

This beaming can be understood more simply if we make some approximations. The
photon's initial energy is small, so for illustrative purposes, we may neglect it
entirely. Then the kinematics of the process are the same as if the incoming
photon did not exist; instead, it looks as if the electron has simply emitted
a photon. The electron behaves almost like a massless particle at these energies,
so the process looks almost like the splitting of one massless particle into two.
The only way that energy and momentum conservation can be satisfied during such
a process is if all three momenta are aligned. When the electron's mass and the
photon's initial energy are taken into account, small deviations from perfect
collinearity are allowed, but these are unimportant. Just as in the synchrotron
case, the vast majority of the radiation is beamed into an extremely narrow pencil
of angles around the direction of the electron's velocity.

In the electron's rest frame, the usual relationship between the initial and final
photon energies $\epsilon_{i}$ and $\epsilon_{f}$ is
\begin{equation}
\epsilon^{e}_{f}=\frac{\epsilon^{e}_{i}}{1+\frac{\epsilon^{e}_{i}}{m}
(1-\cos\theta^{e})}.
\end{equation}
Transformed into the observers
frame, this is (in the $v\rightarrow 1$ limit),
\begin{equation}
\epsilon^{o}_{f}=\epsilon^{o}_{i}\gamma^{2}\frac{(1-\cos\psi_{i}^{o})(1+\cos
\theta^{e}\cos\psi^{e}_{i})}{1+\gamma\frac{\epsilon_{i}^{o}}{m}(1-\cos\psi_{i}^{o})
(1-\cos\theta^{o})}.
\end{equation}
The Lorentz factor $\gamma$ is that of the electron. Depending on the value of
$\gamma\epsilon_{i}^{o}/m$, the final photon energy may be ${\cal O}(\gamma)$ to
${\cal O}(\gamma^{2})$ larger than its initial energy. If $\gamma\epsilon_{i}^{o}/m
\approx 0.1$, for example, then the maximum energy that can be carried off by
the photon is $\left(\epsilon_{f}^{o}\right)_{\max}\approx 0.4\gamma m$.
So the IC process
allows some electrons to transfer sizable fractions of their energy to photons,
and observed IC photon energies range up to $\sim100$ TeV.

All these calculations have been done in the absence of Lorentz violation. With
Lorentz violation, the two most important results continue to hold. The radiation is
strongly
beamed along the direction of the velocity, and the highest-energy IC photons
can carry off significant fractions of the electrons' energies. Of course, with
Lorentz violation, we must be careful to distinguish between quantities such as
$\gamma$ and $E/m$, but this is not a serious complication.

\section{Analysis of Spectral Data}

\label{sec-analysis}

The raw data from which our bounds must be derived are the spectra of high-energy
photon sources. These spectral profiles provide information about the emitting
electrons' energy and velocity distributions. The particular quantities we want to
extract are $\gamma_{\max}$ and $E_{\max}$, so that we may use (\ref{eq-2sidebound})
to bound $c$. Lower limits on these two quantities can be extracted in a fairly
robust fashion, although for $\gamma_{\max}$, the analysis can be a bit tricky.

We shall review here how the values of $\gamma_{\max}$ and $E_{\max}$ are determined.
The synchrotron part of the spectrum tells us about $\gamma_{\max}$. The
highest-energy photons, which arise from IC scattering, give a lower
bound on $E_{\max}$. The bounds are arrived at in entirely different ways, and
they can be derived completely independently even for a single source.

\subsection{Extracting $\gamma_{\max}$}

As it turns out, the analysis required to find $\gamma_{\max}$ is fairly involved.
From (\ref{eq-nuc}), we can infer that $\gamma_{\max}\propto\sqrt{\nu_{c}/\left|
\vec{B}\,\right|}$. The constant of proportionality depends on the sine of the
pitch angle; however, the $\gamma_{\max}$ inferred from assuming $\sin\theta=1$
is always less than the true $\gamma_{\max}$.
The cutoff at $\nu_{c}$ is an obvious feature of synchrotron spectra, and a lower
bound on this cutoff frequency can easily be obtained from any spectrum that
clearly has a synchrotron origin.
However, the tricky part is calculating the strength of
the magnetic field.

The magnetic field is generally taken to be the minimum energy field which can
generate the low-frequency part of the synchrotron spectrum. This can be estimated
fairly accurately by the following calculation. For illustrative purposes, we shall
neglect most numerical constants, but these would of course be retained in a more
detailed calculation. More details are given for the case of a truncated power
law electron spectrum in~\cite{ref-pacholczyk}, and for a broken power law
in~\cite{ref-jester}. Since only the
relatively low-energy part of the spectrum is needed to calculate the magnetic field
strength, we may also ignore the Lorentz violation.

Let $N(E)$ be the energy distribution of the electrons. The total synchrotron
luminosity of these electrons between the energies $E_{1}$ and $E_{2}$ is
\begin{eqnarray}
L & = & \int_{E_{1}}^{E_{2}}dE\, \frac{2e^{4}}{3m^{2}}\gamma^{2}\vec{B}^{2}\sin^{2}
\theta N(E) \\
& \propto & \frac{e^{4}}{m^{4}}\vec{B}^{2}\int_{E_{1}}^{E_{2}}dE\, E^{2}N(E).
\end{eqnarray}
The total energy of the electrons is
\begin{equation}
E_{e}=\int_{E_{1}}^{E_{2}}dE\, EN(E).
\end{equation}
Eliminating the electron distribution between these two and neglecting the
dependence on the lower limit of integration $E_{1}$ gives
\begin{equation}
\label{eq-Ee}
E_{e}\propto\frac{m^{4}}{e^{4}}\frac{L}{\vec{B}^{2}}E_{2}^{-1}.
\end{equation}
Neglecting the dependence on $E_{1}$ is not necessarily a good approximation. In
fact, it is possible for the integrals
to be dominated by the energy range near $E_{1}$.
In that case, the dependence of $E_{e}$ would be on $E_{1}^{-1}$ instead of
$E_{2}^{-1}$, but the results of our calculation would be qualitatively unchanged.
Of course, a great deal more care would be required if we actually wanted to
extract information about a real source, and the full behavior of the integrals
over the entire range of energies would need to be taken into account.

We may replace the energy dependence of (\ref{eq-Ee}) with a frequency dependence.
Since a given electron emits most of its energy around the critical frequency
$\nu_{c}$, we may replace the upper limit on the energy $E_{2}$ with an upper
limit on frequency, and this upper limit $\nu_{2}$ is precisely the critical
frequency for an electron of energy $E_{2}$. So $E_{e}$ finally depends on
\begin{equation}
E_{e}\propto\frac{m^{11/2}}{e^{7/2}}L\left|\vec{B}\,\right|^{-3/2}\nu_{2}^{-1/2}.
\end{equation}

We must also determine the energy that is contained in other particles and in the
magnetic field. We assume that the energy in positively charged particles is
proportional to the energy contained in the electrons. The constant of
proportionality $k$ can potentially range from $k\approx1$ if the positive particles
are positrons, to $k\approx2000$ if the particles are protons accelerated to the
same speeds as electrons. However, although $k$ may cover a wide range, the final
results will depend quite weakly on its value. A change in $k$ by a factor of 100
will change $\gamma_{\max}$ by less than a factor of 2.

The total energy in particles is proportional to $(1+k)L\left|\vec{B}\,\right|
^{-3/2}$, and the energy in the magnetic field is easy to calculate. It is simply
proportional
to $\vec{B}^{2}V$, where $V$ is the volume of the region containing the field.
So the total energy is
\begin{equation}
E_{{\rm tot}}=C\left(\frac{m^{11/2}}{e^{7/2}\nu_{2}^{1/2}}\right)
(1+k)L\left|\vec{B}\,\right|^{-3/2}+\left(\frac{1}{8\pi}\right)\vec{B}^{2}V,
\end{equation}
where $C$ is a numerical constant we have neglected.
We may either minimize this as a function of the magnetic field, or choose
$\left|\vec{B}\,\right|$ according to equipartition, so that the energy in the
magnetic field is equal to the total energy of the particles. Numerically, the
difference between these two methods is negligible, and the dependence of 
the field strength on the other quantities is
\begin{equation}
\left|\vec{B}\,\right|\propto\frac{m^{11/7}}{e}(1+k)^{2/7}\nu_{2}^{-1/7}
V^{-2/7}L^{2/7}.
\end{equation}
This depends very weakly on most of the parameters that must be fitted from the
observed spectrum. Therefore, the value of the field can be determined quite
robustly. By considering only a limited range of frequencies, up to some $\nu_{2}$,
the relevant parameters can be determined just from the lower-energy part of the
spectrum, making the inferred value of $\left|\vec{B}\,\right|$ independent of
the value of $\nu_{c}$ with which it must be combined to give the final value
of $\gamma_{\max}$. Ultimately, $\gamma_{\max}$ depends on at most the
seventh root of any fit parameter (except $\nu_{c}$) that might be in error, and
so its value is quite robust. The limited impact of any possible errors is
discussed at length in~\cite{ref-jester}.

\subsection{Extracting $E_{\max}$}

Placing a lower bound on $E_{\max}$ is much simpler. Models of a source's structure
can yield information about the energies of the electrons doing the emitting.
In general, the models typically require maximum electron energies that are several
times larger
than the highest observed photon energies, because IC scattering events do not
transfer all of the high-energy electrons' energies to the photons. However, to
get a more robust bound, we may take $E_{\max}$ to be the highest actually
observed photon energy; this conservative estimate
usually differs from a model-derived bound by less than an order of magnitude.
The only input we require from a model is that the source's
$\gamma$-ray emission is well described by the IC process.

Choosing $E_{\max}$ in this way
ensures that Lorentz-violating distortions of the energy-momentum relation at
higher than observed energies are not a problem. If the electrons' energy-momentum
relation became significantly Lorentz-violating at the same scale as the highest
particle energies, then the model results might be inaccurate, because they
assume an unmodified electron dispersion relation up to arbitrarily large energies.
However, the maximum observed photon energy is an absolute lower bound on the
electrons' highest energies, independent of whether or not there is Lorentz
violation.

As already stated, the importance of the models is that they can tell us whether
the high-energy end of a source's emission spectrum fits with the hypothesis that
IC scattering is the source of the radiation. There certainly are
astrophysical sources whose spectra are not understood; such sources absolutely
cannot be used to derive bounds on $c$. The quantity $E_{\max}$ is the maximum
energy of subluminal electrons in a source. We can only infer that the electron
involved in a particular IC event is moving more slowly than light if we know that
all the
electrons in a source have subluminal speeds. The signature of superluminal
electrons would be a radiation spectrum that does not fit any known mechanism.
Faster-than-light motion would definitely
represent new physics, so we cannot predict
with any assurance what the radiation from electrons with speeds greater than one
would look like. However, we do expect on very general grounds that they should
radiate energy extremely quickly. For charged particles with 
superluminal speeds, the rate of synchrotron emission diverges if the radiation
reaction force is neglected. There will also be vacuum Cerenkov radiation.
It is conceivable that the poorly understood spectra of some extremely high energy
sources may actually be evidence of superluminal electron motion, and it is
precisely this reason that only well understood IC sources can be used to place
bounds on $E_{\max}$.

\section{Experimental Results and Constraints on $c$}

\label{sec-data}

There are many sources for which measurements of $\gamma_{\max}$ and $E_{\max}$ have
already been made. The sources with the largest
observed values of these quantities will give the best bounds on the
Lorentz-violating coefficients, and we have gleaned the data necessary for setting
optimal bounds on $c$ from the existing observational literature. There are a number
of model-derived values of $\gamma_{\max}$ in the literature, and we have located
seven that are numerically large enough to contribute usefully to the bounds on $c$.
Unfortunately, even for many well-studied sources, there are no
reliable published values of $\gamma_{\max}$, and it might be a worthwhile future
undertaking to determine $\gamma_{\max}$ for more of these sources.

Because the evaluation of $E_{\max}$ is more straightforward, we have found more
useful $E_{\max}$ values than $\gamma_{\max}$ values.
Many of the best measurements of IC radiation presently available come
from the H.E.S.S. telescope in Namibia. It is the excellent sensitivity of this
device that makes many the limits on the IC side possible. However, the sky coverage
of this device is limited, which is a drawback.

Bounds on SME coefficients are generally given in a sun-centered
celestial equatorial coordinate frame~\cite{ref-bluhm4}. Right ascension and
declination constitute a system of polar coordinates for this same reference frame.
However, for parameterizing a quantity such as $c$, it is necessary to introduce
Cartesian coordinates. The origin of the coordinates is at the center of the sun.
The $Z$-axis points along the direction of the Earth's rotation, and the $X$-axis
points toward the vernal equinox
point on the celestial sphere. (That is, the $X$-direction
is the direction from the Earth to the sun at the occurrence of the vernal equinox,
so at the time of this equinox, the Earth lies along the negative $X$-axis.) The
$Y$-direction is chosen according to the right hand rule. The Earth's orbit is
inclined by approximately $23^{\circ}$ from the $XY$-plane. Although it is
unimportant here, the origin of time ($T=0$) is conventionally taken to be at the
vernal equinox in the year 2000.

Since both synchrotron and inverse Compton radiation are strongly beamed along the
direction of an electron's motion, when we observe this radiation from a given
source, we are observing emissions from electrons moving in the source-to-Earth
direction at ultrarelativistic speeds. This gives the correct $\hat{e}$ to
appear in (\ref{eq-2sidebound}).
In terms of the right ascension $\alpha$ and declination $\delta$, the components
of $\hat{e}$ are $\hat{e}_{X}=-\cos\delta\cos\alpha$,
$\hat{e}_{Y}=-\cos\delta\sin\alpha$, and $\hat{e}_{Z}=-\sin\delta$; the minus
signs come from the fact that $\hat{e}$ is the direction for the source to the
Earth, not vice versa.
It is important to note that the velocity of the
source as a whole will not enter into our calculations in any way; we are
looking at the velocities of individual electrons, and the bulk motion of the source
is irrelevant. Of course however, for distant extragalactic sources, cosmological red
shifts would need to be taken into account.

There are nine components of $c$ that can be bounded with the astrophysical
data---the three
$c_{0j}$ and the six-component symmetric part of $c_{jk}$.
Each of the inequalities derived from (\ref{eq-2sidebound}) generally couples all
nine of the coefficients in a
nontrivial way. These bounds may be fairly awkward.
However, the coupled bounds may be translated into bounds on the
separate coefficients by means of linear programming.
The linear program produces absolute bounds on each coefficient; these values
are the largest and smallest that a given coefficient can be under any circumstances.
Of course, there
are also additional correlations, as it is not generally possible for several of the
coefficients to take on their extreme values simultaneously.

To constrain the nine Lorentz-violating coefficients to lie within a bounded
region of the parameter space requires at least ten inequalities. If these
inequalities are all derived from measurements of $\gamma_{\max}$ or $E_{\max}$,
then at least nine sources must be used. Together, the $\gamma_{\max}$ and
$E_{\max}$ bounds for a single source constrain the allowed parameters to the region
between to parallel hyperplanes; nine such hyperplane pairs would be needed to
produce a completely bounded region.

\begin{table}
\begin{center}
\begin{tabular}{|l|c|c|c|c|c|}
\hline
Emission source & $\hat{e}_{X}$ & $\hat{e}_{Y}$ & $\hat{e}_{Z}$ &
$\gamma_{\max}$ & $E_{\max}/m$ \\
\hline
3C 273 & 0.99 & 0.13 & $-0.04$ &
$3\times10^{7}$\cite{ref-roser} & $2\times10^{5}$\cite{ref-roser} \\
Centaurus A & 0.68 & 0.27 & 0.68 & $2\times 10^{8}$\cite{ref-kataoka} & - \\
Crab nebula & $-0.10$ & $-0.92$ & $-0.37$ & $3\times
10^{9}$\cite{ref-aharonian1} & $2\times 10^{8}$\cite{ref-tanimori,ref-aharonian1} \\
G 0.9+0.1 & 0.05 & 0.88 & 0.47 &
- & $10^{7}$\cite{ref-aharonian9} \\
G 12.82-0.02 & $-0.06$ & 0.95 & 0.29 &
- & $5\times 10^{7}$\cite{ref-aharonian6} \\
G 18.0-0.7 & $-0.11$ & 0.97 & 0.24 &
- & $7\times 10^{7}$\cite{ref-aharonian8,ref-aharonian11} \\
G 347.3-0.5 & 0.16 & 0.75 & 0.64 &
$3\times 10^{7}$\cite{ref-ellison} & $2\times 10^{7}$\cite{ref-aharonian3} \\
MSH 15-52 & 0.34 & 0.38 & 0.86 &
- & $8\times 10^{7}$\cite{ref-aharonian4} \\
Mkn 421 & 0.76  & $-0.19$  & $-0.62$ &
- & $3\times 10^{7}$\cite{ref-albert,ref-aharonian7} \\
Mkn 501 & 0.22 &0.74 & $-0.64$ &
- & $4\times 10^{7}$\cite{ref-aharonian10} \\
PSR B1259-63 & 0.42 & 0.12 & 0.90 &
- & $6\times 10^{6}$\cite{ref-aharonian5} \\
RCW 86 & 0.35 & 0.30 & 0.89 &
$10^{8}$\cite{ref-rho} & - \\
SNR 1006 AD & 0.52  & 0.53 & 0.67 &
$2\times 10^7$\cite{ref-allen} & $7\times 10^{6}$\cite{ref-allen} \\
Vela SNR & 0.44 & $-0.55$ & 0.71 & $3\times 10^{8}$\cite{ref-mangano} &
$1.3\times 10^{8}$\cite{ref-aharonian2} \\
\hline
\end{tabular}
\caption{
\label{table-combined}
Parameters for the astrophysical sources that we shall use to constrain $c$.
References are given for each value of
$\gamma_{\max}$ or $E_{\max}$.}
\end{center}
\end{table}

Table~\ref{table-combined} lists the data for fourteen sources for which good
measurements are available. We have added several more sources to the list that was
used in~\cite{ref-altschul6}. These include sources for which good data have only
very recently been released, as well as additional extragalactic sources, such as the
blazars Markarian 421 and Markarian 501.  The additional sources allow for
significant improvements in the bounds on some of the $c$ coefficients.

In addition to the astrophysical bounds, we have included some
other comparable bounds in the linear program. This improves the resolution for
the individual coefficients significantly, because the inequalities derived just from
the astrophysical bounds do not complement one-another optimally. If only the
astrophysical bounds are considered, the
nine-dimensional polytope region within which the $c$ coefficients must lie turns
out to be fairly elongated.
Individual $c^{\nu\mu}$ coefficients are allowed to take on values significantly
larger than the typical $\gamma_{\max}^{-2}$ values, because of possible large
cancellations with other coefficients. However, since bounds derived
from laboratory experiments have completely different forms, they are excellent
complements to the astrophysical constraints and improve the situation markedly.

The additional bounds that we use come from optical resonator tests. These tests are
usually
used to place bounds on the parameters of the SME photon sector. However, the
same experiments may be used to place bounds on the electron $c$
coefficients~\cite{ref-muller2}. The key realization is that electronic Lorentz
violations will modify the structure of a crystalline resonator, and this effect
can be worked out systematically, provided Lorentz violations for nucleons can be
safely neglected.
These experiments then yield measurements of $c_{(XY)}$, 
$c_{(XZ)}$, $c_{(YZ)}$, and $c_{XX}-c_{YY}$. The measured values of these
coefficients in~\cite{ref-muller2} are at roughly the $10^{-15}$ level, with
standard errors of the same magnitude.
\{Note that~\cite{ref-muller2} uses a different convention for
symmetrizing the $c_{jk}$ coefficients, so that the quoted values of $c_{(jk)}$ in
that paper are smaller by a factor of two.\}
There is no compelling evidence from these measurements that any
of the coefficients are nonzero.  Therefore, we shall use these results to derive
bounds that are symmetric about zero. If the measured value and error for a
particular $c_{(jk)}$ is $\alpha\pm\beta$, we shall treat this as a bound
$|c_{(jk)}|<|\alpha|+2\beta$, which is valid at at least the $2\sigma$
level. The resulting bounds on $|c_{(XY)}|$ and 
$|c_{(XZ)}|$ are about $3\times 10^{-15}$; for $|c_{(YZ)}|$ and $|c_{XX}-c_{YY}|$,
they are slightly more stringent, at better than a $2.5\times 10^{-15}$ level.

\begin{table}
\begin{center}
\begin{tabular}{|c|c|c|}
\hline
$c_{\mu\nu}$ & Maximum & Minimum \\
\hline
$c_{XX}$ & $5\times 10^{-15}$ & $-3\times 10^{-15}$ \\
$c_{YY}$ & $2.5\times 10^{-15}$ & $-7\times 10^{-16}$ \\
$c_{ZZ}$ & $2.5\times 10^{-15}$ & $-1.6\times 10^{-15}$ \\
$c_{(XY)}$ & $3\times 10^{-15}$ & $-3\times 10^{-15}$ \\
$c_{(YZ)}$ & $3\times 10^{-15}$ & $-3\times 10^{-15}$ \\
$c_{(YZ)}$ & $1.8\times 10^{-15}$ & $-2.5\times 10^{-15}$ \\
$c_{0X}$ & $4\times 10^{-15}$ & $-7\times 10^{-15}$ \\
$c_{0Y}$ & $1.5\times 10^{-15}$ & $-5\times 10^{-16}$ \\
$c_{0Z}$ & $2\times 10^{-17}$ & $-4\times 10^{-17}$ \\
\hline
\end{tabular}
\caption{
\label{table-separate}
Independent bounds on the components of $c$. The astrophysical data do not improve
on the results of~\cite{ref-muller2} for $c_{(XY)}$ and $c_{(XZ)}$. The lower
limit on $c_{(YZ)}$ is also controlled solely by the data from~\cite{ref-muller2};
however, the upper limit has been improved by adding the astrophysical information.}
\end{center}
\end{table}

The output of the linear program is given in table~\ref{table-separate}.
Since these bounds represent the
absolute maximum and minimum values that are possible for each coefficient, they
are not always as tight numerically as the raw bounds.
We observe that the cryogenic resonator bounds
are still the best for $|c_{(XY)}|$ and  $|c_{(XZ)}|$; the lower limit on
$c_{(YZ)}$ also comes solely from~\cite{ref-muller2}. However, the bounds
on almost all the other coefficients are comparable to or better than the resonator
bounds, and there are some significant improvements over the results presented
in~\cite{ref-altschul6} as well. The bounds on
$c_{0Z}$ are especially strong, while in laboratory experiments,
boost invariance violation coefficients such as $c_{0j}$ are typically harder to
constrain. This shows the advantage of deriving bounds from emissions by
relativistic sources.


%
%
%
%
%
%
%

\section{Conclusion}

\label{sec-concl}

The various bounds in table~\ref{table-separate} are generally similar in their
orders of magnitude, and this observation is actually quite interesting.
The methods we have used could not actually detect a nonzero $c^{\nu\mu}$ directly.
Instead, the signature of Lorentz violation
in high-energy astrophysical sources would be emission spectra that
could not be modeled
by conventional radiation mechanisms. If the calculated bounds on certain
Lorentz-violating coefficients were significantly weaker than others, that would be
an indication that those coefficients might actually be nonzero. However, we see no
indications that any particular components of $c$ are more likely to be
nonzero than others.

The bounds that can be derived by combining the astrophysical and resonator data
are in some cases many orders of magnitude better than the bounds on the same
coefficients that could be extracted by other methods. Doppler shift measurements
can only constrain the $c_{0j}$ coefficients to be less than about $10^{-2}$.
So the improvements here are by more than
twelve orders of magnitude (fourteen orders for $c_{0Z}$). Moreover, with better
sky coverage and more sensitive detectors, even greater precision ought to be
possible.

However, there are some fundamental limits to the how accurate these astrophysical
measurements can be.
There are different ways to search for Lorentz violation. Laboratory tests generally
involve extremely high precision measurements. One may
then compare the results of these measurements when they are
made in different reference frames. The rotation and revolution of
the earth naturally provide a selections of observation frames with different
orientations and velocities. Some experiments also utilize measurement apparatuses
that rotate in the laboratory or beams of particles moving with substantial
velocities in the lab frame. In all these kinds of scenarios,
experimental accuracy is the most important limitation to setting tight bounds.
Questions relating to metrology and long-term experimental stability can be very
important, and the experiments can be extremely demanding technically.
However, as the accuracy with which measurements can be made improves, the
bounds on Lorentz violation will always improve accordingly. Bounds on Lorentz
violations can usually be smaller than the errors involved in the absolute
measurement
of a quantity, because Lorentz violations possess characteristic signatures
(such as sidereal variations) whose existence can often be very strongly excluded.

Astrophysical tests of Lorentz symmetry are rather different. The astrophysical
bounds on $k_{F}$ and $k_{AF}$ come from searches for photon birefringence. The
data sets involved are noisy; the sources are irregular, and the detectors are
not always especially precise. However, extremely tight bounds may be derived by
taking advantage of cosmological distances. A tiny effect on the propagation of
photons will be magnified by an incredibly long line of sight to the source. It is
this line of sight that is the most important limitation on the precision of these
bounds, and so these bounds cannot generally be improved by systematic improvements
the way laboratory bounds can.

The bounds derived from astrophysical synchrotron and IC sources are
similar. In this case, it is not very large astrophysical distances that we are
utilizing, but very high astrophysical energies. At these energies, miniscule
Lorentz-violating terms can lead to obvious changes in the observed spectra.
However, again these bounds are not subject to perpetual improvement, however much
the systematics of our measurements improve. Better
sky coverage and more accurate detectors surely will allow us to improve these
bounds somewhat, but there is an ultimate limit that we cannot pass beyond.
This limit
is set by the actual maximum energies of electrons in these sources. Above some
energy, there will simply exist too few electrons for us to observe their radiation,
no matter how sensitive our detectors become. So other
laboratory-based methods of constraining the
$c$ coefficients, perhaps relying on precision measurements of atomic transitions
at optical frequencies, may ultimately produce more accurate results than these
astronomical methods.

However, at present, the astrophysical bounds we have derived are the best currently
available for most
of the $c^{\nu\mu}$ coefficients. The bounds on all the coefficients are at about
the $10^{-15}$ level or a little better.  As better experimental data become
available, the bounds will continue to improve, and these results are further
clear demonstrations of the usefulness of astrophysical data in constraining
violations of Lorentz invariance.

\section*{Acknowledgments}
The author is grateful to V. A. Kosteleck\'{y} and Q. G. Bailey for helpful
discussions.
This work is supported in part by funds provided by the U. S.
Department of Energy (D.O.E.) under cooperative research agreement
DE-FG02-91ER40661.

\end{document}